\documentclass[twocolumn,pra,showpacs,preprintnumbers,superscriptaddress,amsmath,amssymb,10pt,aps]{revtex4}

\usepackage{mathptmx}
\usepackage{subfigure}
\usepackage{psfrag,graphicx}
\usepackage{dcolumn}
\usepackage{amsmath,amssymb}
\usepackage{bm}
\usepackage{color}
\usepackage{latexsym}
\usepackage{epstopdf}
\usepackage{color}
\usepackage[english]{babel}
\usepackage{latexsym}
\usepackage{psfrag,graphicx}
\usepackage{subfigure}
\usepackage{amsmath}
\usepackage{amssymb}
\usepackage{amsfonts}
\usepackage{bm}
\usepackage{natbib}
\usepackage{epstopdf}
\usepackage{appendix}

\definecolor{mygrey}{gray}{0.35}
\definecolor{myblue}{rgb}{0.2,0.2,0.8}
\definecolor{myzard}{cmyk}{0,0,0.05,0}
\definecolor{mywhite}{rgb}{1,1,1}
\definecolor{mywhite}{rgb}{1,1,1}
\definecolor{myred}{rgb}{1,0.,0.3}

\usepackage[colorlinks=true,citecolor=myblue,linkcolor=myred]{hyperref}

\def\ba{\begin{align}}
\def\enda{\end{align}}
\def\bi{\begin{itemize}}
\def\ei{\end{itemize}}

\def\be{\begin{equation}}
\def\ee{\end{equation}}
\def\bea{\begin{eqnarray}}
\def\eea{\end{eqnarray}}
\def\bse{\begin{subequations}}
\def\ese{\end{subequations}}

\begin{document}
\title{Quantum parameter estimation of nonlinear coupling in trilinear Hamiltonian with trapped ions }
\author{Peter A. Ivanov}
\affiliation{Department of Physics, St. Kliment Ohridski University of Sofia, James Bourchier 5 blvd, 1164 Sofia, Bulgaria}

\begin{abstract}
I propose an efficient method for measuring non-linear coupling between the collective axial breathing mode and the radial rocking mode induced by the mutual Coulomb repulsion in linear ion crystal. The quantum sensing technique is based on the laser induced coupling between one of the vibrational modes and the internal ion's spin states which allows to estimate the non-linear coupling either by measuring the phonon probability distribution or directly be observing the Ramsey-type oscillations of the ion spin states. I show that due to the presence of non-linear phonon coupling the off-resonance interaction between the ion spin states and the axial breathing mode leads to spin-dependent phonon squeezing of the radial rocking mode. Thus the non-linear coupling can be estimated by measuring population distribution of the motional squeezed state. Furthermore, I show that the off-resonance interaction between the spin and the radial rocking mode creates a spin-dependent beam splitter operation between the two vibrational modes. Thus, the parameter estimation can be carried out by detecting the ion spin populations. Finally, I show that the measurement uncertainty precision can reach the Heisenberg limit by using an entangled states between the two collective modes.
\end{abstract}

\maketitle

\section{Introduction}
High-precision quantum sensors are of great importance in testing fundamental physics and quantum technologies \cite{Degen2017,Pezze2018}. One of the most promising quantum platform with application in quantum sensing is the system of laser-cooled trapped ions, which allows excellent control over the internal and motional degrees of freedom \cite{Blatt2008}. Examples include high-precision frequency measurement \cite{Meyer2001,Leibfried2004,Chwalla2009}, sensing of weak electric fields and forces \cite{Biercuk2010,Shaniv2017,Wolf2019,Burd2019,Gilmore2021}, and ultrasensitive magnetometer \cite{Kotler2014,Baumgart2016,Ruster2017}. Up to harmonic approximation of the mutual Coulomb repulsion between the ions the collective vibrational modes are decoupled. However, under a given resonance conditions of the trap frequencies the cross-Kerr nonlinear coupling between the collective modes becomes significant \cite{Marquet2003,Nie2009}. The precise determination of such nonlinear coupling is essential for the realization of non-Gaussian gates with continuous variables \cite{Lau2016} as well as pushing the fidelity of two-qubit quantum gates towards the fault-tolerant threshold. A small frequency shift due to the non-linear phonon coupling has been observed in two ion crystal \cite{Roos2008}. Recently, a frequency shift of the motional mode that is proportional to the number of phonons in another motional mode was observed \cite{Ding,Ding2017}.

In this work I propose sensing scheme for measuring small phonon non-linear coupling, which use a spin-motional coupling to transfer the relevant information of the parameter we wish to estimate into the phonon state distribution or directly into the spin-degrees of freedom. In the first sensing scheme the spin states of the ion is coupled to the axial breathing mode via off-resonance dipolar interaction described by the quantum Rabi model \cite{Lv2018}. I show that dispersive spin-phonon interaction suppresses a phonon exchange between the two modes and causes a spin-dependent \emph{motional squeezing} of the radial rocking mode, with magnitude proportional to the non-linear phonon coupling. Thus, the parameter estimation can be carried out by detecting phonon state probability distribution \cite{Lechner2016,Kirkova2021}. I quantify the sensitivity of the parameter estimation in terms of Fisher information and show that the measurements of the radial rocking mode phonon distribution saturates the quantum Cram\'er-Rao bound. In the second sensing scheme the spin states are coupled with the radial rocking vibrational mode. In that case the dispersive spin-phonon interaction induces a spin-dependent \emph{beam-splitter operation} between the two vibrational modes. I show that for given initial motional states the non-linear phonon coupling can be detected by observing the Ramsey-type oscillations of the ion spin states. Moreover, the sensitivity of the parameter estimation can be enhanced by increasing the number of phonons. I show that oscillation frequency of the signal is amplified by a factor proportional to the number of phonons so that the Heisenberg limit of precision can be achieved.

The paper is organized as follows: In Sec. \ref{NLCC} I provide the general background of the Coulomb mediated cross-Kerr nonlinearity between the collective vibration modes. In particular I discuss the non-linear coupling between the axial breathing mode and the radial rocking mode in a linear ion crystal with two ions. In Sec. \ref{QP} I discuss the laser-ion interaction in the presence of non-linear phonon coupling. I show that the off-resonance bichromatic laser interaction can be used to map the relevant information of the non-linear coupling into the motional Fock state probabilities. In Sec. \ref{QE} I consider the sensitivity of the parameter estimation using the language of the quantum parameter estimation theory. Finally, the conclusions are presented in Sec. \ref{C}.

\section{Non-linear Coulomb phonon coupling}\label{NLCC}

Consider a system of $N$ ions with charge $e$ and mass $m$ confined in a radio-frequency Paul trap along the $z$ axis with trap frequencies $\omega_{\alpha}$ ($\alpha=x,y,z$). The potential energy of the system consists of an effective harmonic potential and mutual Coulomb repulsion \cite{Leibfried2003,James1998},
\begin{equation}
\hat{V}=\frac{m}{2}\sum_{\alpha=x,y,z}\sum_{k=1}^{N}\omega^{2}_{\alpha}\hat{r}^{2}_{\alpha,k}+\sum_{k>k^{\prime}}^{N}\frac{e^{2}}{4\pi\epsilon_{0}|\hat{\vec{r}}_{k}-\hat{\vec{r}}_{k^{\prime}}|}.\label{V}
\end{equation}
Here $\epsilon_{0}$ is the permitivity of free space and $\hat{\vec{r}}_{k}=(\hat{r}_{x,k},\hat{r}_{y,k},\hat{r}_{z,k})$ is the position radius vector operator of ion $k$.
For low temperature and sufficiently strong radial trap frequencies $\omega_{x,y}\gg\omega_{z}$, the ions are arranged in a linear configuration and occupy equilibrium position
$\hat{\vec{r}}_{k}=(0,0,lu_{k})$ where $l=(e^{2}/4\pi\epsilon_{0}m\omega_{z})^{1/3}$ is the length scale which characterizes the average distance between the ions and $u_{k}$ is the dimensionless equilibrium position. The displacement of each ion from its equilibrium position is denoted with $\delta \hat{r}_{\alpha,k}$. Making a Taylor expansion of the external trapping potential and the mutual Coulomb repulsion between the ions  around the equilibrium position $u_{k}^{0}$ the Hamiltonian for the ion's vibration becomes
\begin{eqnarray}
\hat{H}_{\rm v}&=&\frac{1}{2m}\sum_{\alpha=x,y,z}\sum_{k=1}^{N}\hat{p}^{2}_{\alpha,k}+\frac{m}{2}\sum_{\alpha=x,y,z}\sum_{k,k^{\prime}=1}^{N}A^{\alpha}_{k,k^{\prime}}
\delta \hat{r}_{\alpha,k}\delta \hat{r}_{\alpha,k^{\prime}}\notag\\
&&+\hat{H}_{\rm NL}.\label{H}
\end{eqnarray}
Within the harmonic approximation the elements of $(N\times N)$ real and symmetric matrix $A^{\alpha}_{k,k^{\prime}}$ are given by
\begin{equation}
A_{k,k^{\prime}}^{\alpha}=\left\{\begin{array}{c}
\beta^{2}_{\alpha}+\sum_{r=1,r\neq k^{\prime}}^{N}\frac{b_{\alpha}}{|u_{k^{\prime}}-u_{r}|^{3}},\ (k=k^{\prime}), \\
-\frac{b_{\alpha}}{|u_{k}-u_{k^{\prime}}|^{3}},\ (k\neq
k^{\prime}),
\end{array}\right.
\end{equation}
where $\beta_{\alpha}=\omega_{\alpha}/\omega_{z}$, $b_{x}=b_{y}=-1$ and $b_{z}=2$. Up to the quadratic terms in the displacement the ion's vibrations along the three spatial directions are decoupled. The collective modes can be determined by solving $\sum_{k^{\prime}=1}^{N}A_{k,k^{\prime}}^{\alpha}b^{\alpha(p)}_{k^{\prime}}=\gamma_{\alpha,p}b^{\alpha(p)}_{k}$, ($p=1,2,\ldots,N$), where $\gamma_{\alpha,p}$ are the eigenvalues and $\bold{b}^{\alpha(p)}$ are the corresponding eigenvectors. Indeed, one can introduce the normal mode coordinates operators $\hat{Q}_{\alpha,p}=\sum_{k=1}^{N}b_{k}^{\alpha(p)}\delta \hat{r}_{\alpha,k}$ and the corresponding conjugate momentum operators $\hat{P}_{\alpha,p}=\sum_{k=1}^{N}b_{k}^{\alpha(p)}\hat{p}_{\alpha,k}$ which transform the quadratic part of Hamiltonian (\ref{H}) into canonical form, $\hat{H}_{\rm quad}=\sum_{\alpha}\sum_{p}\{(\hat{P}^{2}_{\alpha,p}/2m)+m\omega^{2}_{\alpha}\hat{Q}^{2}_{\alpha,p}/2\}$. Hence, within the harmonic approximation the collective ion vibration can be described in terms of a set of independent harmonic oscillators with characteristic normal mode frequencies given by $\omega_{\alpha,p}=\sqrt{\gamma_{\alpha,p}}\omega_{\alpha}$.

The higher order terms in the Taylor expansion of the potential (\ref{V}) however give rise to non-linear Coulomb mediated couplings between the collective modes, which is described by the term $\hat{H}_{\rm NL}$. The non-linear coupling between the collective modes vary with the number of ions and can be found in \cite{Marquet2003}. For concreteness we assume ion string with two ions $N=2$ where we have
\begin{equation}
\hat{H}_{\rm NL}=-\frac{m\omega^{2}_{z}}{2^{5/6}l}\hat{Q}_{z,2}\{2\hat{Q}^{2}_{z,2}-3\hat{Q}^{2}_{x,2}-3\hat{Q}^{2}_{y,2}\}+O(\hat{Q}^{4}_{\alpha,p}),
\end{equation}
The Hamiltonian $\hat{H}_{\rm NL}$ describes a non-linear interaction between the axial breathing mode with frequency $\omega_{b}=\sqrt{3}\omega_{z}$ and the two radial rocking modes with frequencies $\omega_{{\rm rock},q}=\sqrt{\omega^{2}_{q}-\omega^{2}_{z}}$ ($q=x,y$). Furthermore, one can defined phonon creation and annihilation operators $\hat{a}^{\dag}_{\alpha,p}$ and $\hat{a}_{\alpha,p}$ of the $p$th collective phonon mode so that $\hat{Q}_{\alpha,p}=\sqrt{\hbar/2m\omega_{\alpha,p}}(\hat{a}^{\dag}_{\alpha,p}+\hat{a}_{\alpha,p})$ and respectively $\hat{P}_{\alpha,p}=i\sqrt{\hbar m\omega_{\alpha,p}/2}(\hat{a}^{\dag}_{\alpha,p}-\hat{a}_{\alpha,p})$.

Imposing the resonance condition $\omega_{b}\approx2\omega_{r}$ and neglecting the fast oscillating terms we obtain
\begin{equation}
\hat{H}_{\rm v}=\hbar\omega_{b}\hat{a}^{\dag}_{b}\hat{a}_{b}+\hbar\omega_{r}\hat{a}^{\dag}_{r}\hat{a}_{r}+\hbar\lambda(\hat{a}_{b}\hat{a}^{\dag2}_{r}
+\hat{a}^{\dag}_{b}\hat{a}^{2}_{r})+\hat{H}^{\prime}.\label{Hvv}
\end{equation}
Here we have denoted $\omega_{z,b}=\omega_{b}$, $\omega_{x,{\rm rock}}=\omega_{r}$ and respectively $\hat{a}_{z,2}=\hat{a}_{b}$ and $\hat{a}_{x,2}=\hat{a}_{r}$. The third term in (\ref{Hvv}) describes the trilinear interaction between the two collective modes with coupling strength $\lambda=\omega_{b}(z_{b}/2^{5/6}l)$ with $z_{b}=\sqrt{\hbar/2m\omega_{b}}$. The Hamiltonian $\hat{H}^{\prime}$ describes the fast rotating terms which can be neglected as long as $\omega_{b}\gg\lambda$, $|\omega_{y,{\rm rock}}-\omega_{b}|\gg\lambda$.

\section{Quantum Probe}\label{QP}

Our probe system consists of a two level system with a metastable internal states $\left|\uparrow\right\rangle$ and $\left|\downarrow\right\rangle$ and transition frequency $\omega_{0}$. The interaction-free Hamiltonian is $\hat{H}_{\rm free}=\hat{H}_{\rm v}+\frac{\hbar\omega_{0}}{2}\sigma^{z}_{1}$ where $\sigma^{x,y,z}_{1}$ are the Pauli matrices. We assume that the ion is simultaneously addressed by bichromatic laser fields with a wave-vector $\vec{k}$ along the spatial direction $\alpha=x,y,z$. The Hamiltonian describing the laser-ion interaction, after making the optical rotating-wave approximation, is given by \cite{Wineland1998,Schneider2012,Hafner2008}
\begin{equation}
\hat{H}_{I}(t)=\hbar\Omega\{\sigma^{+}_{1}e^{i k_{\alpha}\delta \hat{r}_{\alpha,1}}(e^{-i\omega_{R}t-i\phi_{R}}+e^{-i\omega_{B}t-i\phi_{B}})+{\rm H.c.}\}\label{HI}
\end{equation}
Here $\Omega$ is the Rabi frequency and $\sigma^{\pm}_{1}$ are the raising and lowering operators for the effective spin system. $\omega_{R}$, $\omega_{B}$ are the laser frequencies and $\phi_{R}$, $\phi_{B}$ are the corresponding laser phases.

\subsubsection{Case 1}

Consider that the bichromatic laser fields propagate along the axial trap axis, $\alpha=z$. Assume that the frequency beat notes are set to $\omega^{(1)}_{R}=\omega_{0}-\omega_{b}+\omega$ and $\omega^{(1)}_{B}=\omega_{0}+\omega_{b}-\omega$ which drives the simultaneous red- and blue-sideband transitions between the internal ion states and the axial breathing mode with detuning $\omega$. We also assume that the two trap frequencies satisfy the condition $\omega_{b}=2\omega_{{\rm rock}}+\omega$.  Within the Lamb-Dicke limit  the Hamiltonian (\ref{HI}) in the rotating-frame with respect to $\hat{U}_{1}(t)=e^{-i\omega_{0}t\sigma^{z}_{1}/2-i(\omega_{b}-\omega)t\hat{a}^{\dag}_{b}\hat{a}_{b}-i\omega_{\rm rock}t\hat{a}^{\dag}_{r}\hat{a}_{r}}$, so that $\hat{H}^{(1)}_{I}=\hat{U}_{1}^{\dag}(\hat{H}_{\rm v}+\hat{H}_{I}(t))\hat{U}_{1}-i\hbar \hat{U}_{1}^{\dag}\partial_{t}\hat{U}_{1}$ is given by
\begin{eqnarray}
&&\hat{H}^{(1)}_{I}=\hat{H}^{(1)}_{0}+\hat{H}^{(1)}_{\rm sb},\quad \hat{H}^{(1)}_{0}=\hbar\omega \hat{a}^{\dag}_{b}\hat{a}_{b},\notag\\
&&\hat{H}^{(1)}_{\rm sb}=\hbar g_{b}\sigma^{x}_{1}(\hat{a}^{\dag}_{b}+\hat{a}_{b})+\hbar\lambda(\hat{a}_{b}\hat{a}^{\dag2}_{r}+\hat{a}^{\dag}_{b}\hat{a}^{2}_{r}),\label{HInew}
\end{eqnarray}
where $g_{b}=\Omega\eta_{z}$ is the spin-phonon coupling, $\eta_{z}=k_{z}z_{b}$ standing for the Lamb-Dicke parameter ($\eta_{z}\ll1$) and we set $\phi_{R}=\phi_{B}=\pi/2$. At the first inspection, evolution under the Hamiltonian $\hat{H}^{(1)}_{I}$ seems complicated. Indeed, the first term in $\hat{H}^{(1)}_{\rm sb}$ is the quantum Rabi Hamiltonian which describes the dipolar coupling between the two-level system and the breathing phonon mode which entangles the spin and motional states, while the second term leads to coherent energy exchange between the two motional modes in which one phonon from the axial breathing mode is converted into a pair of phonons in the radial rocking mode and vice versa. However, as we will discuss below under the condition of high frequency $\omega$ compared to the coupling $g_{b}$ the breathing phonon excitations are suppressed. However, due to the presence of non-linear coupling the spin-phonon interaction causes motional squeezing of the radial rocking mode with magnitude proportional to $\lambda$.
\subsubsection{Case 2}

Alternatively, consider that the laser fields propagate along the radial direction $\alpha=x$. Then, assume that the red- and blue-sidenabd laser frequencies are set close to the rocking mode, $\omega^{(2)}_{R}=\omega_{0}-\omega_{r}+\omega$ and $\omega^{(2)}_{B}=\omega_{0}+\omega_{r}-\omega$ and respectively the trap frequencies satisfy the condition $\omega_{b}=2\omega_{r}-\omega$. Using this and transforming the interaction Hamiltonian (\ref{HI}) in rotating frame with respect to $\hat{U}_{2}(t)=e^{-i\omega_{0}t\sigma^{z}_{1}/2-i(\omega_{b}-\omega)t\hat{a}^{\dag}_{b}\hat{a}_{b}-i(\omega_{r}-\omega)t\hat{a}^{\dag}_{r}\hat{a}_{r}}$ we obtain
\begin{eqnarray}
&&\hat{H}^{(2)}_{I}=\hat{H}^{(2)}_{0}+\hat{H}^{(2)}_{\rm sb},\quad \hat{H}^{(2)}_{0}=\hbar\omega \hat{a}^{\dag}_{b}\hat{a}_{b}+\hbar\omega \hat{a}^{\dag}_{r}\hat{a}_{r},\notag\\
&&\hat{H}^{(2)}_{\rm sb}=\hbar g_{r}\sigma^{x}_{1}(\hat{a}^{\dag}_{r}+\hat{a}_{r})+\hbar\lambda(\hat{a}_{b}\hat{a}^{\dag2}_{r}+\hat{a}^{\dag}_{b}\hat{a}^{2}_{r}),\label{HInew2}
\end{eqnarray}
where $g_{r}=\Omega\eta_{x}$ with $\eta_{x}=k_{x}x_{r}$ and $x_{r}=\sqrt{\hbar/2m\omega_{r}}$. We will consider below regime of weak excitation of the two collective modes. Remarkably, in that case the parameter $\lambda$ can be estimated by observing the Ramsey-type oscillations of the spin populations.
\begin{figure}
\includegraphics[width=0.45\textwidth]{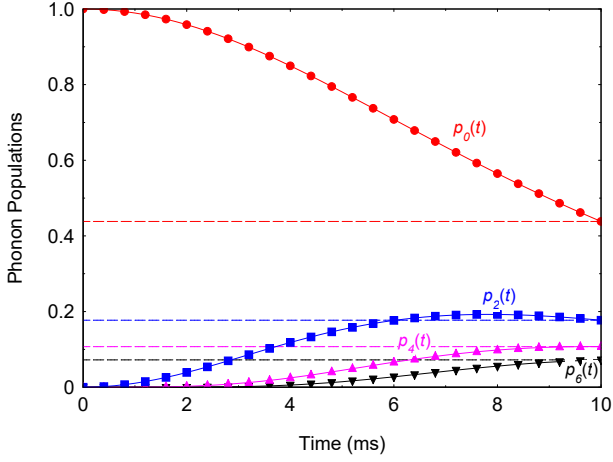}
\caption{(Color online) Fock state probabilities $p_{n_{r}}(t)$ for radial rocking mode as a function of time. We integrate numerically the time-dependent Schr\"odinger equation with Hamiltonian (\ref{HInew}) with initial state $|\psi(0)\rangle=|+\rangle|0_{b}\rangle|0_{r}\rangle$. The parameters are set to $g_{b}/2\pi=3.5$ kHz, $\omega/2\pi=15$ kHz, and $\lambda/2\pi=0.05$ kHz. For comparison we plot the corresponding Fock state probabilities $p_{2n_{r}}(t_{f})=\frac{\tanh^{2n_{r}}(r)}{\cosh(r)}\frac{(2n_{r})!}{2^{2n_{r}}(n_{r}!)^{2}}$ at $t_{f}=10$ ms with squeezing parameter $r=(2g_{b}\lambda/\omega)t_{f}$ (dashed lines).}
\label{fig1}
\end{figure}

\subsection{Adiabatic Elimination}
\subsubsection{General Approach}
In the following we consider weak-coupling regime $g_{s(r)},\lambda\ll \omega$ in which one of the phonon degrees of freedom can be eliminated from the dynamics. This can be carried out by performing the canonical transformation of Hamiltonian (\ref{HInew}) and (\ref{HInew2}) so that $\hat{H}^{(a)}_{\rm eff}=e^{-\hat{S}_{a}}(\hat{H}^{(a)}_{0}+\hat{H}^{(a)}_{\rm sb})e^{\hat{S}_{a}}$ ($a=1,2$), where $\hat{S}_{a}$ is anti-Hermitian operator $\hat{S}_{a}^{\dag}=-\hat{S}_{a}$. Using Baker-Campbell-Hausdorff expression we have
\begin{eqnarray}
\hat{H}^{(a)}_{\rm eff}&=&\hat{H}^{(a)}_{0}+\hat{H}^{(a)}_{\rm sb}+[\hat{H}^{(a)}_{0},\hat{S}_{a}]+[\hat{H}^{(a)}_{\rm sb},\hat{S}_{a}]\notag\\
&&+\frac{1}{2!}[[\hat{H}^{(a)}_{0},\hat{S}_{a}],\hat{S}_{a}]+\frac{1}{2!}[[\hat{H}^{(a)}_{\rm sb},\hat{S}_{a}],\hat{S}_{a}]+\ldots
\end{eqnarray}
The goal is to choose $\hat{S}_{a}$ in a such a way that all terms of order of $g_{s(r)}$ in $\hat{H}^{(a)}_{\rm eff}$ are canceled and the first terms describing the spin-phonon and phonon-phonon interactions are of order of $g_{s(r)}^{2}/\omega$, $g_{s(r)}\lambda/\omega$, and $\lambda^{2}/\omega$. Indeed, one can determine $\hat{S}_{a}$ by the condition $\hat{H}^{(a)}_{\rm sb}+[\hat{H}^{(a)}_{0},\hat{S}_{a}]=0$ which ensures that the terms of order of $g_{s(r)}$ vanish. Hence, the effective Hamiltonian becomes $\hat{H}^{(a)}_{\rm eff}\approx \hat{H}^{(a)}_{0}+\frac{1}{2}[\hat{H}^{(a)}_{\rm sb},\hat{S}_{a}]$. Next, we transform $\hat{S}_{a}$ into interaction picture with respect to $\hat{H}^{(a)}_{0}$ such that $i\hbar\partial_{t}\hat{S}_{a}(t)=[\hat{S}_{a}(t),\hat{H}^{(a)}_{0}]$, thereby $i\hbar\partial_{t}\hat{S}_{a}(t)=\hat{H}_{\rm sb}(t)$ where $\hat{H}^{(a)}_{\rm sb}(t)=e^{i\hat{H}^{(a)}_{0}t/\hbar}\hat{H}^{(a)}_{\rm sb}e^{-i\hat{H}^{(a)}_{0}t/\hbar}$.
\subsection{Case 1}
Using (\ref{HInew}) we find
\begin{equation}
\hat{S}_{1}=\frac{g_{b}}{\omega}\sigma^{x}_{1}(\hat{a}_{b}-\hat{a}_{b}^{\dag})+\frac{\lambda}{\omega}(\hat{a}_{b}\hat{a}_{r}^{\dag 2}-\hat{a}_{b}^{\dag}\hat{a}_{r}^{2})
\end{equation}
and the effective Hamiltonian becomes
\begin{eqnarray}
&&\hat{H}^{(1)}_{\rm eff}=\hbar\omega\hat{a}_{b}^{\dag}\hat{a}_{b}-\hbar\frac{g_{b}\lambda}{\omega}\sigma^{x}_{1}(\hat{a}_{r}^{\dag 2}+\hat{a}_{r}^{2})+\hat{H}_{\rm res},\notag\\
&&\hat{H}_{\rm res}=\hbar\frac{\lambda^{2}}{\omega}\{4\hat{a}^{\dag}_{b}\hat{a}_{b}\left(\hat{a}_{r}^{\dag}\hat{a}_{r}+\frac{1}{2}\right)-\hat{a}_{r}^{\dag2}\hat{a}_{r}^{2}\}.\label{Heff}
\end{eqnarray}
We see that in the leading order the phonon exchange between the two collective modes is suppressed. Moreover, the off-resonance interaction between the spin system and the axial breathing mode induces a spin-dependent motional squeezing of the radial rocking mode with magnitude $g_{b}\lambda /\omega$. Therefore, measuring rocking phonon distribution we can estimate the parameter $\lambda$. The residual term $\hat{H}_{\rm res}$ describes the nonlinear phonon-phonon interaction which however as long as $\lambda\ll\omega$ it can be neglected.

\subsection{Case 2}

Using (\ref{HInew2}) we find
\begin{equation}
\hat{S}_{2}=\frac{g_{r}}{\omega}\sigma^{x}_{2}(\hat{a}_{r}-\hat{a}_{r}^{\dag})-\frac{\lambda}{\omega}(\hat{a}_{b}^{\dag}\hat{a}_{r}^{2}-\hat{a}_{b}\hat{a}_{r}^{\dag2}).
\end{equation}
The effective Hamiltonian becomes
\begin{equation}
\hat{H}^{(2)}_{\rm eff}=\hbar\omega(\hat{a}^{\dag}_{b}\hat{a}_{b}+\hat{a}^{\dag}_{r}\hat{a}_{r})-\hbar\frac{2g_{r}\lambda}{\omega}\sigma^{x}_{1}(\hat{a}^{\dag}_{b}\hat{a}_{r}+\hat{a}_{b}\hat{a}^{\dag}_{r})-\hat{H}_{\rm res}.\label{Heff123}
\end{equation}
The unitary evolution generated by $\hat{H}^{(2)}_{\rm eff}$ is given by the spin-dependent beam-splitter operator. As we will see later on by proper choice of the initial motional state one can map the relevant information on the parameter $\lambda$ to the spin degree of freedom. Therefore, the parameter estimation is carried out by observing the Ramsey-type oscillations of the ion spin states.

\section{Quantum estimation of $\lambda$}\label{QE}
In this section I will provide the background of the quantum estimation theory. Because the parameter $\lambda$ is not a direct observable its value can be estimated only by performing suitable measurements of other experimentally accessible observable. Consider that the information of the parameter $\lambda$ can be acquired for a given probe state and a specific set of measurement outcomes with probability $p_{n}(\lambda)$ with $n=1,2,\ldots,N$. The classical Fisher information (CFI) then is given by \cite{Paris2009}
\begin{figure}
\includegraphics[width=0.45\textwidth]{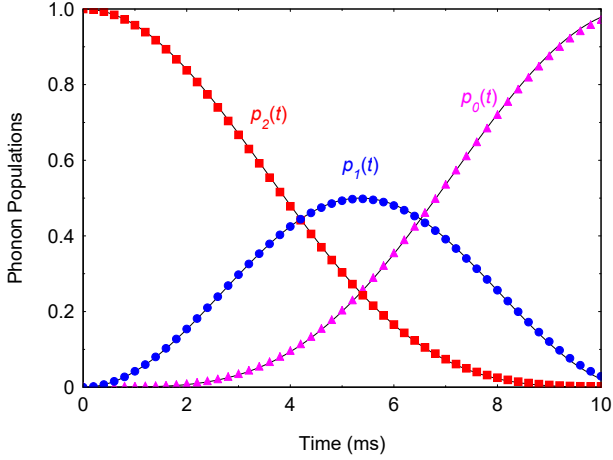}
\caption{(Color online) Fock state probabilities $p_{n_{b}}(t)$ for breathing mode as a function of time. We integrate numerically the time-dependent Schr\"odinger equation with Hamiltonian (\ref{HInew2}) with initial state $|\psi(0)\rangle=|+\rangle|2_{b}\rangle|0_{r}\rangle$. The parameters are set to $g_{r}/2\pi=3.5$ kHz, $\omega/2\pi=45$ kHz, and $\lambda/2\pi=0.15$ kHz. For comparison we plot the corresponding Fock state probabilities $p_{2}(t)=\cos^{4}(\theta t)$, $p_{1}(t)=2\sin^{2}(\theta)\cos^{2}(\theta t)$ and $p_{0}(t)=\sin^{4}(\theta t)$ with $\theta=2g_{r}\lambda/\omega$ (solid lines).}
\label{fig2}
\end{figure}
\begin{equation}
\mathcal{F}_{\rm Cl}(\lambda)=\sum_{n=1}^{N}\frac{1}{p_{n}}\left(\frac{\partial p_{n}}{\partial \lambda}\right)^{2}.\label{Fisher_cl}
\end{equation}
The variance of the parameter estimation is bounded by the classical Cram\'er-Rao inequality
\begin{equation}
\delta\lambda\geq\frac{1}{\sqrt{\nu \mathcal{F}_{\rm Cl}(\lambda)}},
\end{equation}
where $\nu$ is the experimental repetitions. Furthermore, the optimal strategy to estimate the value of $\lambda$ is associated with a privileged observable which maximized the CFI. Indeed, the CFI is upper bounded by $\mathcal{F}_{\rm Cl}\leq \mathcal{F}_{\rm Q}(\lambda)$ where $\mathcal{F}_{\rm Q}(\lambda)={\rm Tr}\left(\hat{\rho}_{\lambda}(t)\hat{\mathcal{L}}_{\lambda}^{2}\right)$ is the quantum Fisher information (QFI). Here $\hat{\rho}_{\lambda}(t)$ is the density operator and $\hat{\mathcal{L}}_{\lambda}$ is the symmetrical logarithmic derivative (SLD) operator, which satisfies the operator equation $2\partial_{\lambda}\hat{\rho}_{\lambda}=\hat{\rho}_{\lambda}\hat{\mathcal{L}}_{\lambda}+\hat{\mathcal{L}}_{\lambda}\hat{\rho}_{\lambda}$ \cite{Paris2009}. Therefore, the ultimate precision in the estimation of the parameter $\lambda$, optimized over all possible measurements, is given by the quantum Cram\'er-Rao bound
\begin{equation}
\delta\lambda\geq\frac{1}{\sqrt{\nu \mathcal{F}_{\rm Q}(\lambda)}}.
\end{equation}
Finally, the optimal measurement basis in which the quantum Cram\'er-Rao bound is saturated is given by the eigenvectors of SLD operator.

\begin{figure}
\includegraphics[width=0.45\textwidth]{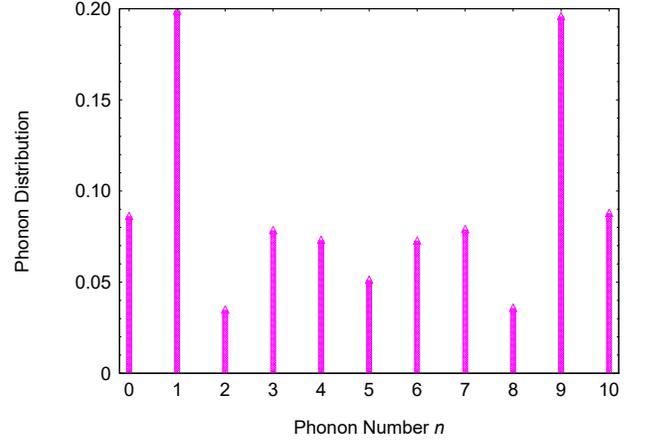}
\caption{(Color online) Phonon distribution of the axial breathing mode at time $t=8$ ms. We integrate numerically the time-dependent Schr\"odinger equation with Hamiltonian (\ref{HInew2}) with initial state $|\psi(0)\rangle=|+\rangle|5_{b}\rangle|5_{r}\rangle$. The parameters are set to $g_{r}/2\pi=3.5$ kHz, $\omega/2\pi=50$ kHz, and $\lambda/2\pi=0.08$ kHz.}
\label{fig3}
\end{figure}

\subsection{Case 1}

Due to the presence of non-linear phonon-phonon coupling the off-resonant interaction between the spin and the axial breathing mode creates squeezing of the radial rocking mode. Indeed, neglecting the residual term in (\ref{Heff}) we arrive in the effective time-evolution dictated by the spin-dependent squeeze operator $\hat{U}_{1}(t)=e^{i\frac{g_{b}\lambda}{\omega}t\sigma^{x}_{1}(\hat{a}^{\dag2}_{r}+\hat{a}^{2}_{r})}$ for mode $r$. Thus detecting the squeezing parameter $r=(2g_{b}\lambda/\omega)t$ we can determine $\lambda$. In Fig. \ref{fig1} we show the exact result for the Fock state probabilities $p_{n_{r}}$ as a function of time where we assume that the system is prepared initially in the state $\hat{\rho}_{\rm in}=|+\rangle\langle+|\otimes\hat{\rho}_{b}\otimes|0_{r}\rangle\langle 0_{r}|$, with $\sigma_{x}|\pm\rangle=\pm|\pm\rangle$ and $\hat{\rho}_{b}$ is the initial density operator for the breathing mode. Note that in the limit $\lambda\ll\omega$ the energy exchange between the two modes is suppressed and the dynamics of the breathing mode is decoupled from the rest of the system. Therefore, the parameter estimation is not affected by the initial phonon state of the breathing mode.

The state at time $t$ is given by $|\psi_{\lambda}\rangle=\hat{S}(\xi)|0_{r}\rangle$ where $\hat{S}(\xi)=e^{\frac{r}{2}(e^{-i\phi}\hat{a}^{2}-e^{i\phi}\hat{a}^{\dag2})}$ being the squeeze operator and $\phi=-\pi/2$. The quantum Fisher information then is given by $\mathcal{F}_{\rm Q}(\lambda)=4\{\langle\partial_{\lambda}\psi_{\lambda}|\partial_{\lambda}\psi_{\lambda}\rangle-|\langle\psi_{\lambda}|\partial_{\lambda}\psi_{\lambda}\rangle|^{2}\}
=2(\partial_{\lambda}r)^{2}$ (see Appendix \ref{QFI_SS}) and optimal estimation precision can be written as
\begin{equation}
\delta\lambda\geq \frac{\omega}{\sqrt{8\nu}g_{b}t}.
\end{equation}
The result indicates that we can improve the sensitivity of the parameter estimation by lowering the ratio $\omega/g_{b}$ but keeping $\omega>g_{b}$.

Finally, it is straightforward to show that the measurements of the phonon state probabilities $p_{n_{r}}$ leads to equality $\mathcal{F}_{\rm Cl}(\lambda)=\mathcal{F}_{\rm Q}(\lambda)$ and hence it saturates the fundamental quantum Cram\'er-Rao bound. The detection of the motional states can be achieved by observing the evolution of the spin states under the influence of a red-sideband interaction \cite{Meekhof1996}. Recently, it was shown that arbitrary phonon state distribution can be measured by using adiabatic transition which map the phonon state probabilities directly into collective spin-excitations \cite{Kirkova2021}.
\begin{figure}
\includegraphics[width=0.45\textwidth]{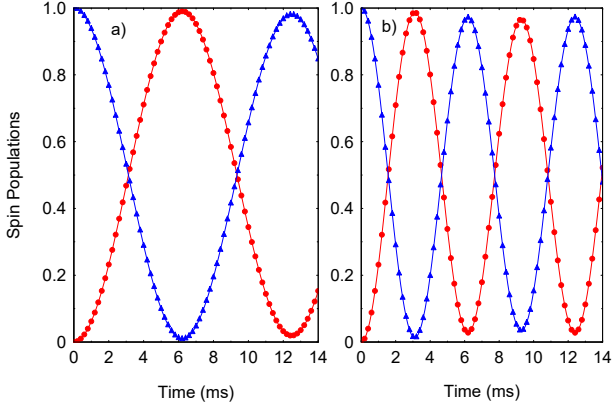}
\caption{(Color online) a) Spin state populations $p_{\uparrow}(t)$ and $p_{\downarrow}(t)$ as a function of time. We integrate numerically the time-dependent Schr\"odinger equation with Hamiltonian (\ref{HInew2}) with initial state $|\psi_{\rm in}\rangle=\left|\downarrow\right\rangle(|1_{b}\rangle|0_r\rangle+|0_b\rangle|1_r\rangle)/\sqrt{2}$. The parameters are set to $g_{r}/2\pi=3.0$ kHz, $\omega/2\pi=60$ kHz, and $\lambda/2\pi=0.4$ kHz. b) The same but with initial state $|\psi_{\rm in}\rangle=\left|\downarrow\right\rangle(|2_{b}\rangle|0_r\rangle+|0_b\rangle|2_r\rangle+\sqrt{2}|1_{b}\rangle|1_{r}\rangle)/2$.}
\label{fig4}
\end{figure}
\subsection{Case 2}
The unitary evolution generated by the spin-phonon interaction with Hamiltonian (\ref{Heff123}) is given by the spin-dependent beam-splitter operator $\hat{U}_{2}(t)=e^{i\theta t\sigma_{1}^{x} (\hat{a}^{\dag}_{b}\hat{a}_{r}+\hat{a}_{b}\hat{a}^{\dag}_{r})}$ with $\theta=2g_{r}\lambda/\omega$. Such a operator is used as an entangles of output fields \cite{Mandel1995,Kim2002}. Recently, a conditional beam-splitter quantum gate between the ion's spin states and the two radial modes of single trapped ion was demonstrated \cite{Gan2020}. As we see the relevant information of the parameter $\lambda$ is encoded in the phase $\theta$ and thus it can be determined by detecting the phonon state populations e.g., either by detecting axial breathing or radial rocking phonon distributions. Consider that the system is initially prepared in the state $|\psi_{\rm in}\rangle=|+\rangle|n_{b}\rangle|0_{r}\rangle$ with $n_{b}$ phonons in the breathing mode and zero phonons in the rocking mode. Then the system evolved into the state $|\psi(t)\rangle=\hat{U}_{2}(t)|\psi_{\rm in}\rangle$ with Fock state probability distribution to observe state with $k_{b}$ phonons in the breathing mode given by
\begin{equation}
p_{k_{b}}(t)=\left( \begin{array}{c} k_{b} \\ n_{b}-k_{b} \end{array} \right)\sin^{2(n_{b}-k_{b})}(\theta t)\cos^{2k_{b}}(\theta t),\label{bs}
\end{equation}
with $k_{b}=0,1,\ldots,n_{b}$. In Fig. \ref{fig2} we compare the analytical expressions for the phonon Fock state probabilities with the exact results for initial number of phonons $n_{b}=2$, where very good agreement is observed. Using (\ref{bs}) we can evaluate the classical Fisher information (\ref{Fisher_cl}). We find $\mathcal{F}_{\rm Cl}(\lambda)=16n_{b}g_{r}^{2}t^{2}/\omega^{2}$. Therefore, the uncertainty in the determination of $\lambda$ is given by
\begin{equation}
\delta\lambda\geq\frac{\omega}{4\sqrt{\nu n_{b}}g_{r}t}.\label{CR2}
\end{equation}
As we see from Eq. (\ref{CR2}) the parameter estimation is improved for higher number of initial number of phonons in the breathing $n_{b}$. This is the standard quantum limit in accuracy for measurement of $\lambda$. Furthermore, we can improve the sensitivity of the parameter estimation by preparing the system initially in the state $|\psi_{\rm in}\rangle=|+\rangle|n_{b}\rangle|n_{r}\rangle$. For simplicity we set $n_{b}=n_{r}=n$. Then the probability to observe $k_{b}$ phonons in the breathing mode is (see, Fig. \ref{fig3})
\begin{eqnarray}
p_{k_{b}}(t)&=&|\sum_{k,l=0}^{n}(-1)^{n-k}\sin^{2n-k-l}(\theta t)\cos^{k+l}(\theta t)\notag\\
&&\times\frac{n!\sqrt{k_{b}!(n-k+l)!}}{k!(n-k)!l!(n-k)!}\delta_{k_{b},n+k-l}|^{2}.\label{population}
\end{eqnarray}
Using (\ref{population}) I find that the classical Fisher information becomes $\mathcal{F}_{\rm Cl}(\lambda)=(32 g_{r}^{2}t^{2}/\omega^{2})n(n+1)$, hence the sensitivity of the parameter estimation is improved at the Heisenberg limit.

Finally, we can map the information of $\lambda$ directly in the spin population. Then the parameter estimation can be carried out by observing the Ramsey-type oscillations the spin state populations. Indeed, let us perform transformation of the phonon operators according to $\hat{a}_{b}=(\hat{c}_{1}+\hat{c}_{2})/\sqrt{2}$ and $\hat{a}_{r}=(\hat{c}_{1}-\hat{c}_{2})/\sqrt{2}$ which gives $\hat{U}_{2}(t)=e^{i\theta t\sigma^{x}_{1}(\hat{c}^{\dag}_{1}\hat{c}_{1}-\hat{c}^{\dag}_{2}\hat{c}_{2})}$. Then assume that the initial state is $|\psi_{\rm in}\rangle=\left|\downarrow\right\rangle(|1_{b}\rangle|0_r\rangle+|0_b\rangle|1_r\rangle)/\sqrt{2}$ which corresponds to $\langle\psi_{\rm in}|\hat{c}^{\dag}_{1}\hat{c}_{1}|\psi_{\rm in}\rangle=1$ and $\langle\psi_{\rm in}|\hat{c}^{\dag}_{2}\hat{c}_{2}|\psi_{\rm in}\rangle=0$. This initial state can be created as follows: after the motional state cooling and optical pumping we begin with state $\left|\downarrow\right\rangle|0_{b}\rangle|0_{r}\rangle$. Then a $\pi/2$ pulse is applied on the blue-sideband transition of the axial breathing mode so that $(\left|\downarrow\right\rangle|0_{b}\rangle+\left|\uparrow\right\rangle|1_{b}\rangle)|0_{r}\rangle$. Finally, a $\pi$ pulse on the carrier transition and subsequent a $\pi$ pulse of the red-sideband transition of the radial rocking mode transform the state into $|\psi_{\rm in}\rangle$.

After the initial state preparation the system evolves for time $t$ according the spin unitary propagator $\hat{U}_{2}(t)=e^{i\theta t\sigma^{x}_{1}}$, so that the spin probabilities becomes $p_{\downarrow}(t)=\cos^{2}(\theta t)$ and $p_{\uparrow}(t)=\sin^{2}(\theta t)$, see Fig. \ref{fig4}(a). Since, the spin evolution depends on the number of phonons in the two modes we can further improve the sensitivity of the parameter estimation. Indeed, assume that the initial state is $|\psi_{\rm in}\rangle=\left|\downarrow\right\rangle(|2_{b}\rangle|0_r\rangle+|0_b\rangle|2_r\rangle+\sqrt{2}|1_{b}\rangle|1_{r}\rangle)/2$ which gives $\langle\psi_{\rm in}|\hat{c}_{1}\hat{c}_{1}|\psi_{\rm in}\rangle=2$ and $\langle\psi_{\rm in}|\hat{c}_{2}\hat{c}_{2}|\psi_{\rm in}\rangle=0$ and hence amplifies the phase by factor of two, see Fig. \ref{fig4}(b). In general, for initial motional state which satisfies the condition $|\psi_{\rm mot}\rangle=\frac{1}{\sqrt{n!}}\left(\frac{\hat{a}^{\dag}_{b}+\hat{a}^{\dag}_{r}}{\sqrt{2}}\right)^{n}|0_{b}\rangle|0_{r}\rangle$ ($n=1,2,\ldots$) we have $p_{\downarrow}(t)=\cos^{2}(n\theta t)$ and $p_{\uparrow}(t)=\sin^{2}(n\theta t)$ and the corresponding classical Fisher information becomes $\mathcal{F}_{\rm Cl}(\lambda)=(16 g_{r}^{2}t^{2}/\omega^{2})n^{2}$ which gives accuracy of the parameter estimation at the Heisenberg limit.

\section{Conclusion}\label{C}

In this work I have proposed quantum sensing schemes for detection of the coupling which quantifies the non-linear interaction between the collective vibrational modes in linear ion crystal. Such an interaction arises due to the high-order terms in the Coulomb repulsion between the trapped ions and which becomes non-negligible under certain resonance condition of the trap frequencies. In a system of two trapped ions this gives rise to an interaction between the axial breathing mode and the radial rocking mode which is described by trilinear Hamiltonian. I have discussed the laser bichromatic interaction between the ion's internal spin states and one of the collective modes in the presence of non-linear phonon interaction. I have shown that the off-resonance dipolar interaction between the breathing mode and the spin described by the quantum Rabi model induces a spin-dependent motional squeezing of the radial rocking mode with magnitude proportional to the non-linear coupling. Hence the parameter estimation can be performed by measuring the phonon state population distribution. Furthermore, I have shown that the off-resonance dipolar interaction between the spin and the radial rocking mode gives rise to a spin-dependent beam-splitter operation between the two collective modes with phase proportional to the non-linear coupling. I have shown that the sensitivity of the parameter estimation which is achieved by measuring the breathing phonon state distribution for initial motional states with $n$ phonons in the two collective modes is enhanced by a factor of $n$. Moreover, I have shown that the information of the non-linear coupling can be extracted by observing the time-evolution of the ion's spin states, which simplify significantly the experimental procedure. Finally, although the proposed quantum sensing technique is focused on the estimation of the coupling characterized the non-linear interaction between two modes, it can be extended to other trilinear Hamiltonians such as $\hat{H}_{\rm NL}=\hbar\lambda(\hat{a}^{\dag}\hat{b}\hat{c}+\hat{a}\hat{b}^{\dag}\hat{c}^{\dag})$ which was experimentally realized in a system of three trapped ions \cite{Ding2018,Maslennikov2019}.

\section*{Acknowledgments}

PAI acknowledges support by the ERyQSenS, Bulgarian Science Fund Grant No. DO02/3.

\appendix
\section{Derivation of the Quantum Fisher Information for squeezed state}\label{QFI_SS}
Here I provide the derivation of the quantum Fisher information for squeezed state. For pure state the quantum Fisher information can be written as $\mathcal{F}_{Q}(\lambda)=4\{\langle\partial_{\lambda}\psi_{\lambda}|\partial_{\lambda}\psi_{\lambda}\rangle-|\langle\psi_{\lambda}|\partial_{\lambda}\psi\rangle|^{2}\}$. For squeezed state we have $|\psi_{\lambda}\rangle=\hat{S}(\xi)|0\rangle$ where $\hat{S}(\xi)=e^{\frac{r}{2}(e^{-i\phi}\hat{a}^{2}-e^{i\phi}\hat{a}^{\dag2})}$ and $r=2g_{b}\lambda t/\omega$, $\phi=-\pi/2$. We have $|\partial_{\lambda}\psi_{\lambda}\rangle=\frac{i}{2}(\partial_{\lambda}r)(a^{\dag2}+a^{2})\hat{S}(\xi)|0\rangle$ and therefore $\langle\psi_{\lambda}|\partial_{\lambda}\psi_{\lambda}\rangle=0$. Using this we find
\begin{eqnarray}
\langle\partial_{\lambda}\psi_{\lambda}|\partial_{\lambda}\psi_{\lambda}\rangle&=&\frac{1}{4}(\partial_{\lambda}r)^{2}\langle0|\hat{S}^{\dag}(\xi)(\hat{a}^{\dag2}+\hat{a}^{2})^{2}
\hat{S}(\xi)|0\rangle\notag\\
&&=\frac{1}{2}(\partial_{\lambda}r)^{2}
\end{eqnarray}
and the quantum Fisher information becomes $\mathcal{F}_{Q}(\lambda)=2(\partial_{\lambda}r)^{2}$.

\end{document}